\documentclass[aps,prl,twocolumn,preprintnumbers]{revtex4}

\usepackage{psfrag,graphicx}
\usepackage{dcolumn}
\usepackage{amsmath,amssymb}
\usepackage{bm}
\usepackage{amsfonts,amssymb,amsmath}        

\newcommand{\SIGMA}{\mbox{\boldmath${\sigma}$}}

\newcommand{\NABLA}{\mbox{\boldmath${\nabla}$}}

\newcommand{\PARTIAL}{\mbox{\boldmath${\partial}$}}
\newcommand{\be}{\begin{equation}}
\newcommand{\ee}{\end{equation}}
\newcommand{\bq}{\begin{eqnarray}}
\newcommand{\eq}{\end{eqnarray}}

\newcommand{\p}{{\bf{p}}}

\newcommand{\1}{1\!\!1}

\begin{document}

\title{Graphene with geometrically induced vorticity}

\author{Jiannis K. \surname{Pachos$^1$}\footnote{Email:
j.k.pachos@leeds.ac.uk},
Michael \surname{Stone$^2$}
and Kristan \surname{Temme$^1$}}
\affiliation{$^1$School of Physics \& Astronomy, University of Leeds,
Leeds LS2 9JT, UK\\
$^2$University of Illinois, Department of Physics 1110 W. Green St. Urbana,
IL 61801 USA}

\date{\today}

\begin{abstract}

At half filling, the electronic structure of graphene can be modelled by a
pair of free two-dimensional Dirac fermions. We explicitly demonstrate that
in the presence of a geometrically induced gauge field, an everywhere-real
Kekul\'e modulation of the hopping matrix elements can correspond to a
non-real Higgs field with non-trivial vorticity. This provides a natural
setting for fractionally charged vortices with localized zero modes. For
fullerene-like molecules we employ the index theorem to demonstrate the
existence of six low-lying states that do not depend strongly on the
Kekul\'e-induced mass gap.

\end{abstract}


\maketitle

Planar graphene and its geometrically related variants offer a rich
environment for exploring interesting physics~\cite{Geim}. The electronic
properties of graphene are well modelled by a simple H{\"u}ckel model of
nearest-neighbor hopping on a two-dimensional honeycomb
lattice~\cite{Wallace,DiVincenzo}. The low energy sector of the resulting
band theory may be described by a pair of two-dimensional Dirac equations.
As a consequence, graphene is expected to exhibit phenomena more familiar in
relativistic quantum theory, such as the Klein Paradox~\cite{Geim_klein}.
Recently, Hou {\em et al.}~\cite{Chamon} demonstrated a relation between
graphene and p-wave superconductors, where fractionally charged vortices can
appear. Energetic considerations guided Jackiw and Pi~\cite{Jackiw} to
extend this model by inserting a gauge potential. Systematic study,
including the demonstration of the fractional statistics of these vortices,
was given in~\cite{Chamon1}. Nevertheless, the question of physically
realizing these systems remained open.

Spherical configurations of graphene are known as fullerenes. The altered
topology requires defects where twelve of the regular carbon hexagons are
replaced by pentagons. The resulting frustration and curvature are accounted
for in the Dirac equation by introducing a chiral gauge field and a spin
connection~\cite{Gonzalez1,Gonzalez2,Semenoff:1984dq,Osipov,Kolesnikov,
Lammert,Cortijo,Neto}. The gauge-field flux (but not the spin-connection
curvature) enters into an index theorem where it is responsible for the six
anomalously low-lying states seen in the spectrum of C$_{60}$ and related
molecules~\cite{Pachos,Pachos1}. Chemists have long surmised that in
fullerenes not all the nearest-neighbor hopping elements are equal. In
numerical calculations with the H{\"u}ckel model it is found that the
molecules can lower their electronic energy by undergoing a small Peierls
distortion --- usually called in this context a Kekul\'e distortion. This
change in the bond lengths introduces a scalar ``Higgs'' field into the
Dirac equation. It has recently been observed, however, that vortices in a
complex-valued Higgs field can bind zero-energy mid-gap
modes~\cite{Chamon,Jackiw,Chamon1,Franz,Ghaemi,Manes}.

Here we demonstrate that the Kekul\'e distortion Higgs field is not a simple
scalar field, but is a section of the non-trivial gauge bundle. This means
that a real-valued modulation of the hopping strengths in a fullerene will
give rise to a complex-valued Higgs field that automatically contains
vortices similar to those of Abrikosov or Nielsen and Olesen. Thus,
fullerenes provide a physical setting where vortices with fractionalized
charge appear naturally. The number of zero modes bound by these vortices is
equal to the number of zero modes required by the index
theorem~\cite{Weinberg,Jackiw_81}. Indeed, they are the {\it same\/} modes.

We begin with the description of a flat sheet of graphene. In the electronic
tight binding (H{\"u}ckel) approximation~\cite{Gonzalez1} electrons hop on a
two-dimensional honeycomb lattice with lattice constant $\alpha$. This
bipartite lattice can be decomposed into two triangular Bravais sub-lattices
$\Lambda_A$ and $\Lambda_B$. The hamiltonian can be written
\be
\hat H = -\sum_{{\bf r}\in \Lambda_A}\sum_{k=1}^3 (t+\delta t_{{\bf r}k} )
a_{\bf r}^\dagger b_{{\bf r}+{\bf s}_k} +{\rm H.c.},
\label{Ham}
\ee
where the vector ${\bf r}$ gives the position of the vertices of the
$\Lambda_A$ lattice and the vectors ${\bf s}_k$, $k=1,2,3$ connect each site
of the $\Lambda_A$ lattice to the three adjacent sites of $\Lambda_B$. The
fermionic operators $a_{\bf r}$ and $b_{{\bf r+s}_k}$ annihilate electrons
in the sub-lattices $\Lambda_A$ and $\Lambda_B$, respectively. At
half-filling, graphene possesses two independent Fermi points at ${\bf
K}_{\pm}=\left(\pm{4 \over 3\sqrt{3} \alpha},0\right)$ where the positive
and negative energy bands touch in conical singularities. For momenta near
these Fermi points we can replace the full hamiltonian with a Dirac
approximation
\be
\hat H = \int \Psi^\dagger ({\bf r})H\Psi({\bf r}) d^2 r
\label{SecQuant}
\ee
where $H$ is the corresponding one-particle Dirac hamiltonian and the spinor
is given by $\Psi({\bf r})^T = (u_a,u_b,v_a,v_b)$ with the $a$, $b$ indices
correspond to the $\Lambda_A$ and $\Lambda_B$ sublattices and $u$, $v$,
correspond to the two Fermi points. For the particular case of the Kekul\'e
distortion given by
\be
\delta t_{{\bf r},k} = \frac{1}{3} \Phi({\bf r})
e^{i{\bf K}_{+}\cdot{\bf s}_k}e^{i{\bf ({\bf K}_+-{\bf K}_-)}
\cdot{\bf r}} + {\rm c.c.},
\ee
the Dirac hamiltonian takes the form~\cite{Chamon}
\be
H=\left(
\begin{array}{cc}
\SIGMA^*\cdot \p & \sigma_1 \Phi\\
\sigma_1 \Phi^* & -\SIGMA \cdot \p \end{array}
\right)
\label{Ham1}
\ee
where $\p = -i \PARTIAL$, and we have taken an overall constant to unity.
Here, $\SIGMA=(\sigma_1,\sigma_2)$ are the two-by-two Pauli matrices.
Transformations of the form $\sigma_i\otimes\1$ act on the Fermi point
components $(u,v)$, while transformations of the form $\1\otimes
\sigma_j$ act on the sublattice indices $(a,b)$. Here we take $\Phi$ to
be initially real everywhere.

We wish to compactify the Kekul\'e distorted graphene sheet into a surface
with the topology of a sphere. To do this it helps to make a change of
basis. We exchange the role of A and B at the ${\bf K_+}$ Fermi point and
rotate the reference frame at the ${\bf K_-}$ point by $\pi$ angle so that
it coincides with the frame at ${\bf K_+}$. These two transformations are
effected by conjugating with the ${\rm SU}(4)$ matrix
$\sigma_1\oplus\sigma_3$. To introduce curvature, we observe that the
Kekul\'e distortion leaves every third hexagon with no double bonds. We
select one of these hexagons and, starting from its geometrical center,
excise a wedge of opening angle $\pi/3$, the cuts passing through the
centers of two lines of bonds. We then reconnect the dangling bonds to form
a seam. This operation leaves us with a conical curvature singularity
centered on a pentagon, but does not cause a dislocation in the pattern of
double bonds. It does, however, introduce frustration in the electron wave
functions. This is because the identity of the A and B lattices is
interchanged across the seam, and because an electron with wavefunction
located near one Dirac point sees itself across the seam as a wavefunction
belonging to the other Dirac point. To make the spinors continuous across
the cut we therefore introduce a gauge-twist transformation
\be
\Psi \rightarrow U\Psi, \,\,\,\text{with}\,\,\, U= \exp\Big[ -i\int (e{\bf
a}+{\bf q})\cdot d{\bf r}\Big]
\ee
when the spinor is transported around the apex of the cone. Here ${\bf
a}=\mbox{\boldmath$A$}(\sigma_2\otimes \1)$ is the non-abelian field with
circulation around the pentagonal plaquette $e\oint {\bf a}\cdot d{\bf r} =
{\pi/ 2}(\sigma_2\otimes \1)$ and ${\bf q}$ is the spin connection with
circulation $\oint {\bf q} \cdot d{\bf r} = -\pi/6(\1\otimes\sigma_3)$.
Taking into account that $U^\dagger\partial_i U =-ieA_i
(\sigma_2\otimes\1)-iq_i$ we have that the momentum operator becomes ${\bf
p}\rightarrow -i({\NABLA} -ie\mbox{\boldmath$A$})$ with $\NABLA =\PARTIAL
-i{\bf q}$. We can diagonalize the gauge field and simplify the Dirac
matrices by the rotation $(e^{-i\sigma_1\pi/4}\otimes\1)(\1\oplus\sigma_3)$
giving finally
\be
H_A=\left(
\begin{array}{cc}
-ie^i_\mu \sigma_i ({\NABLA} -ie\mbox{\boldmath$A$}) &\Phi\\
\Phi^* & ie^i_\mu\sigma_i
({\NABLA} +ie\mbox{\boldmath$A$}) \end{array}
\right)
\label{Ham5}
\ee
where the zweibeins, $e^i_\mu$ are introduced to make the Hamiltonian
covariant in the induced curved space with metric $g_{\mu
\nu}=e_\mu^ie_\nu^i\eta_{ij}$, where $\eta_{ij}$ is the flat metric.
Note that, due to the geometric distortion, the initially real $\Phi$ has
now acquired a phase factor of the form $\exp(i2e\int A)$.

\begin{center}
\begin{figure}[ht]
\resizebox{\linewidth}{!}
{\includegraphics{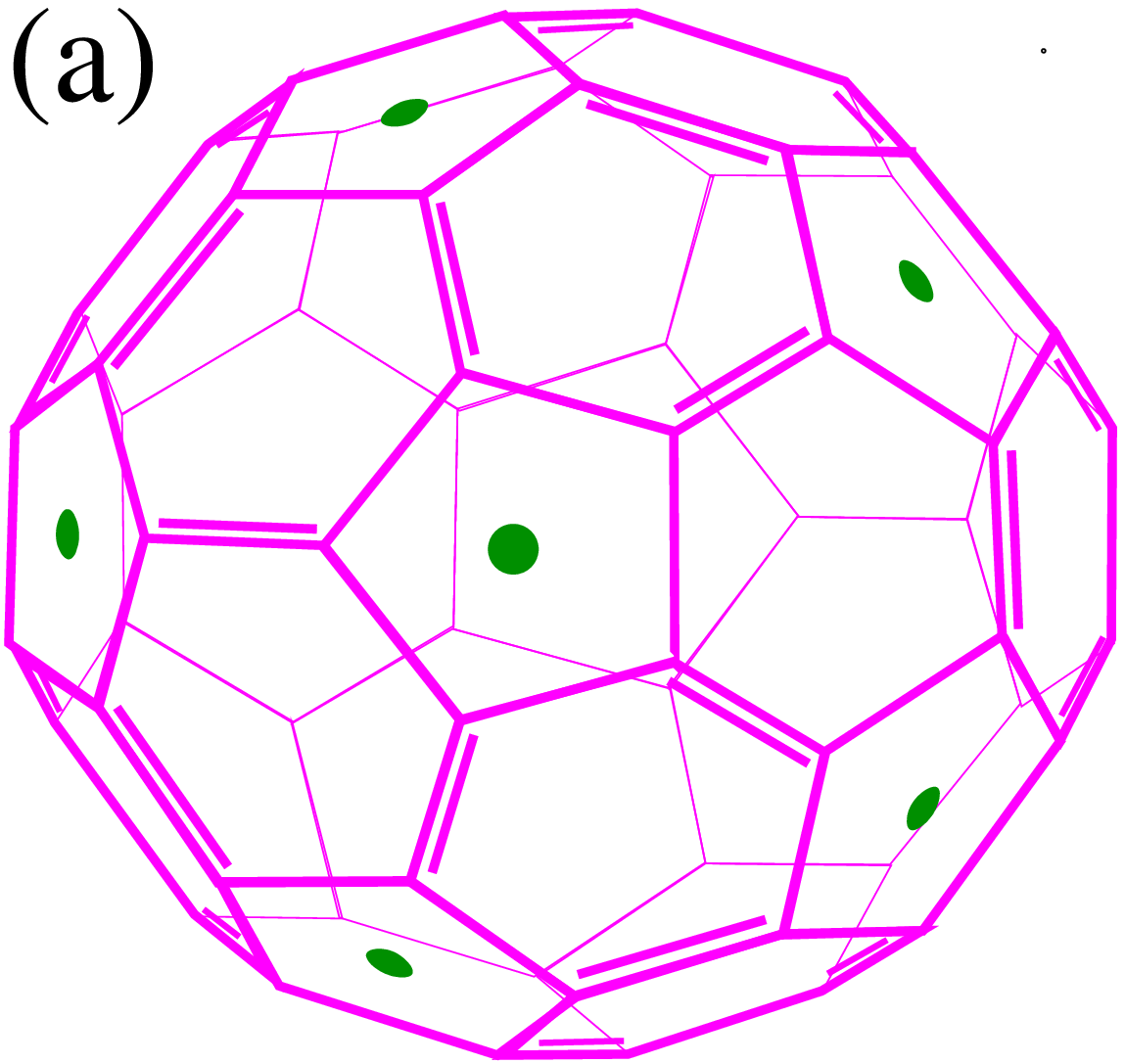}
\hspace{0.5cm}
\includegraphics{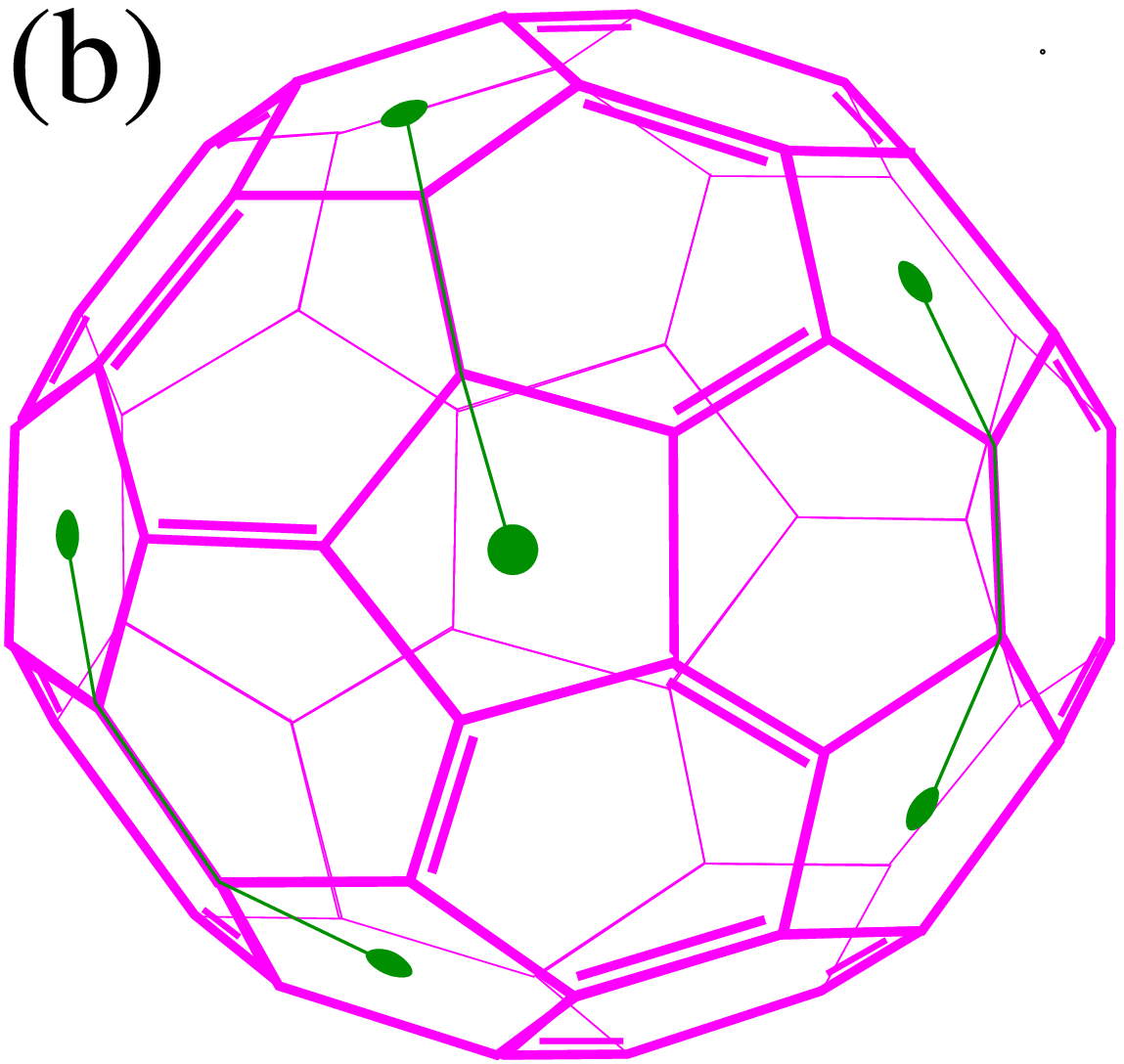}
\hspace{0.5cm}
\includegraphics{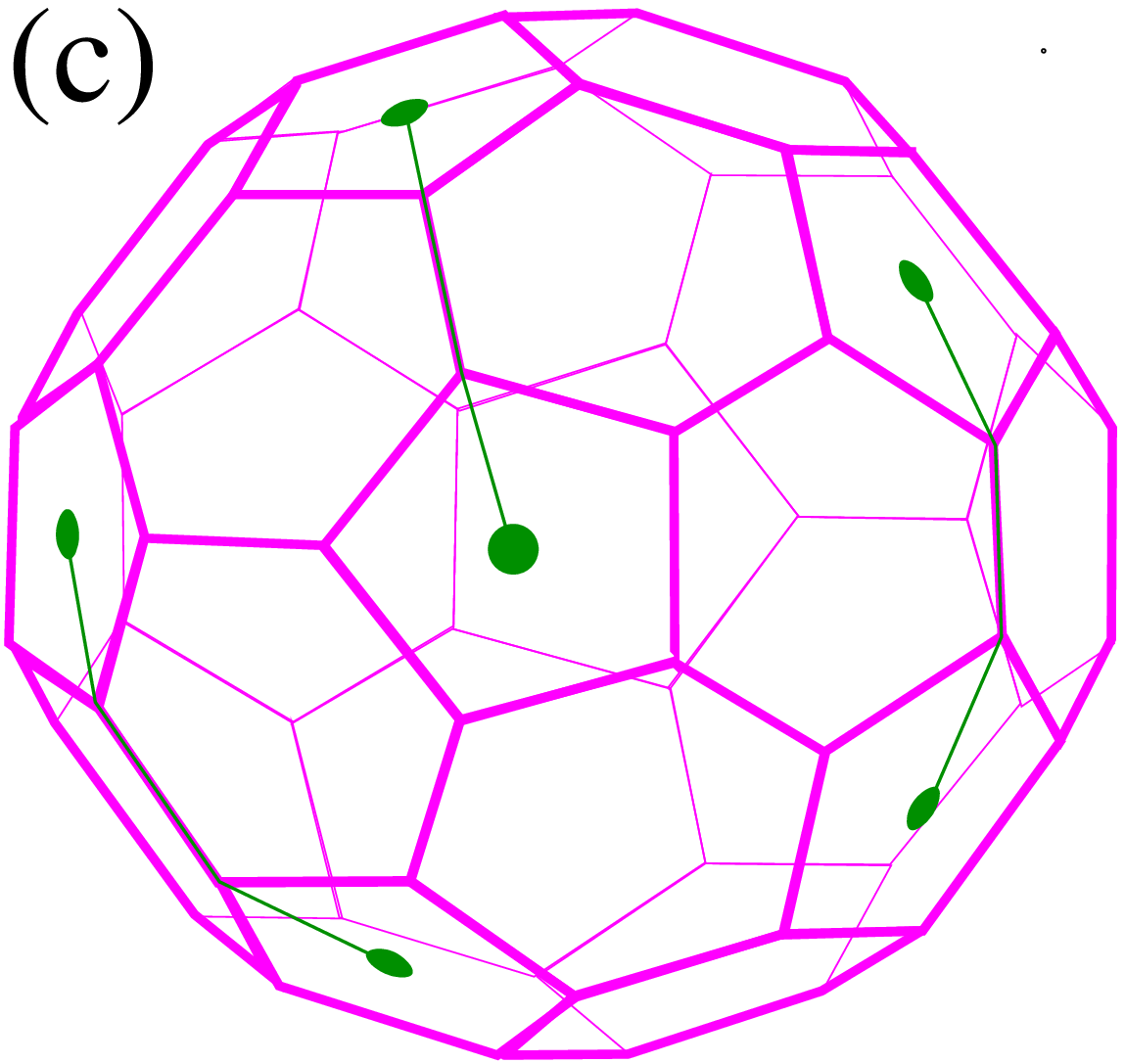}
 }
\caption{\label{C60Kekule} The coupling configuration of the C60
molecule, where vortices reside on the pentagons. (a) The Kekul\'e
distortion. (b) Cuts between vortices are introduced by replacing a double
bond with single ones. (c) An enlargement of the vortices is introduced by
removing all double bonds connected to the pentagons. For C60 this removes
all double bonds.}
\end{figure}
\end{center}

Hamiltonian (\ref{Ham5}) corresponds to a spinor field defined on a curved
manifold coupled to a chiral gauge potential. We observe that the gauge
invariant Dirac operator, $H_A$, that we have derived coincides with the one
recently written down by Jackiw and Pi~\cite{Jackiw}. In our case the
pentagon is threaded by a {\it quarter unit\/} of gauge flux, and that the
phase of the scalar field $\Phi$ winds through only $\pi$ as we circle the
defect. Each pentagon is therefore a rather singular ``half-vortex''. To
make the Higgs field smooth, we would need to join pairs of vortices and
switch off the Kekul\'e distortion along a line joining them (see
Figure~\ref{C60Kekule}(a,b)). We will discuss the significance of this issue
later.

We constructed our hamiltonian (\ref{Ham1}) for the case of a single conical
deformation. In order to make a surface with the topology of a sphere we
formally need to introduce twelve pentagonal defects. Can we extend our
surgical construction globally? It is known that we obtain a consistent
assignment of double and single bonds composing a dislocation-free Kekul\'e
structure (a {\it Fries\/} structure) if and only if the fullerene belongs
to the family C$_{60+6k}$, $k=0,1,\ldots$ that is obtained from a parent
C$_{20+2k}$ molecule by ``leapfrog'' inflation~\cite{leapfrog1}. This is
exactly the family of molecules we obtain by the previous process, where
each of the twelve pentagonal deformations is surrounded by hexagons each
possessing three double bonds. The total gauge flux through the twelve
pentagons is $3\cdot 2\pi$, and the total winding number of the Higgs field
is six, composed of twelve half-vortices. This can be equivalently described
by a monopole sitting inside the molecule providing the net vortex
flux~\cite{Gonzalez1,Gonzalez2}.

At this point it is convenient to study the transformation properties of the
one-particle Dirac hamiltonian $H_A$. First, note that the different
chiralities in the Dirac hamiltonian correspond to normalized solutions,
which have their support on the different triangular sublattices of the
honeycomb lattice. By comparison with $H$ we deduce that the Dirac
hamiltonian $H_A$ acts on the vector $(u_b,u_a,v_a,v_b)$. For convenience we
change basis so that the spinor is given by $(u_a,v_a,u_b,v_b)$. Then one
has $\Gamma_5 H_A\Gamma_5 =-H_A$ where $\Gamma_5=\sigma_3\otimes \1$. This
transformation maps positive energy eigenvectors of $H_A$ to negative ones,
i.e. $\Gamma_5\Psi_E({\bf r}) =\Psi_{-E}({\bf r})$, while zero energy modes
are left unpaired. This property is known as {\em sublattice symmetry} and
it gives rise to charge fractionalization and the emergence of zero modes
when the scalar field has non-trivial vorticity~\cite{Chamon}. Furthermore,
by defining the operator $\Sigma_1 =\1\otimes \sigma_1$ we can show that
$\Sigma_1 H_A\Sigma_1= H_A^*$. As a consequence~\cite{Gurarie} the
eigenfunctions of $H_A$ satisfy $\Sigma_1\Psi =\Psi^*$ and in spinor
components $u_a^*=v_a$ and $u_b^*=v_b$. This symmetry is known as the {\em
time reversal symmetry} and it is present due to the choice of real hopping
elements in hamiltonian (\ref{Ham}).

We wish to apply the index theorem of Jackiw and Rossi~\cite{Jackiw_81}, and
E.~Weinberg~\cite{Weinberg} to our problem. This theorem was first
introduced to study zero modes in a relativistic analogue of superconducting
vortices which is exactly the case here. As we shall see the number of zero
modes provided from the index theorem can be computed by {\it either\/}
counting the number of gauge field flux-units {\it or\/} by counting the net
winding number of the Higgs field
--- the two numbers necessarily being equal because of the topology of the
gauge bundle.

The sublattice symmetry dictates that the non-zero energy eigenfunctions
come in $\pm E$ pairs, whilst zero-energy eigenfunctions (zero modes) can be
chosen to be eigenvectors of $\Gamma_5$. Thus, we can arrange the zero modes
in terms of their chirality. Suppose that there are $n_+$ zero modes with
$\Gamma_5$ eigenvalue $+1$ and $n_-$ with eigenvalue $-1$. Weinberg
shows~\cite{Weinberg} that the ${\rm Index}(H_{\rm Dirac})
\stackrel{\rm def}{=} n_+-n_-$ is given by
\bq
{\rm Index}&&\!\!\!\!\!\!\!\!\!\!\!\!(H_{\rm Dirac})=
\nonumber \\&=& \frac e \pi\int_\Omega Fd^2r +
\frac 1{4\pi i} \oint_{\partial \Omega} d{\bf r} \cdot
\frac{\Phi^* {\bf D} \Phi- \Phi{\bf D} \Phi^*}{|\Phi|^2}
\nonumber\\
&=& \frac 1{2\pi} \oint_{\partial \Omega} d{\bf r}\cdot \PARTIAL{\rm Arg\,}
\Phi.
\eq
Here
$$
{\bf D} \Phi = (\PARTIAL-2ie{\bf A})\Phi,\quad {\bf D}
\Phi ^*= (\PARTIAL+2ie{\bf A})\Phi^*,
$$
and the ${\partial \Omega}$ integral is to be taken over a contour
surrounding all the vortices. It is possible to use the consistency
equation, $(\PARTIAL-2ie{\bf A})\Phi=0$, of the scalar field that follows
from $A_i({\bf r}) = \partial_i\phi({\bf r})$. Then, it is easy to see that
the index can be written in terms of the gauge field
\be
{\rm Index}(H_{\rm Dirac}) = {e\over \pi}  \int_{\Omega}Fd^2r.
\ee
From this form of the index of $H_{\rm Dirac}$ we deduce that one can take
the scalar field continuously to zero, $\Phi\rightarrow 0$, without changing
the number of zero modes.

As derived by Weinberg, the index theorem applies to the case of open
boundary conditions. To apply the theorem to a spherical fullerene with
hamiltonian $H_A$ requires the following step. We recognize that there is a
net gauge-field flux through the sphere, and so the gauge and associated
Higgs bundle are non-trivially twisted. We must therefore introduce two
hemispherical patches and sew them together with a gauge-transformation that
identifies $\PARTIAL - ie{\bf A}$ in the upper hemisphere with $e^{ie\chi}
(\PARTIAL-ie{\bf A})e^{-ie\chi} $ in the lower, and similarly identifies
$\Phi$ in the upper hemisphere with $e^{2ie\chi}\Phi $ in the lower. The
phase $e\chi$ in the ${\rm U}(1)$ group element $e^{ie\chi}$ will wind
through $\int_{S^2}eF=6\pi$ as we encircle the common boundary of the
hemispheres. It is the twisting of the Higgs field that allows us to have a
net number (six) of Higgs field vortices. If the field $\Phi$ were an
ordinary scalar instead of a section of a twisted bundle the net vortex
number would necessarily be zero. Thus, the index theorem demonstrates that
the mass term appearing due to the Kekul\'e distortion does not actually
destroy the zero modes.

\begin{center}
\begin{figure}[h]
\resizebox{\linewidth}{!}
{\includegraphics{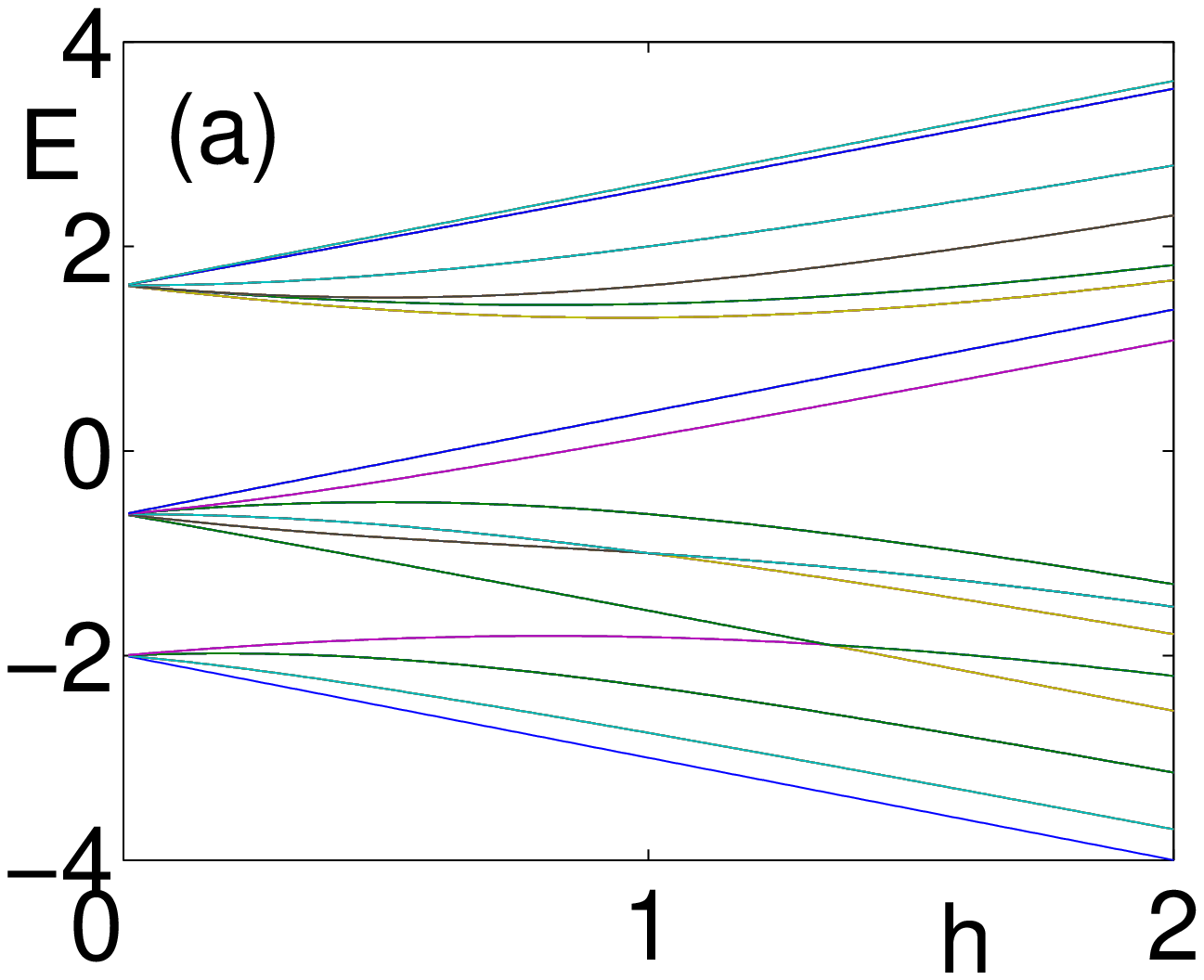}
\includegraphics{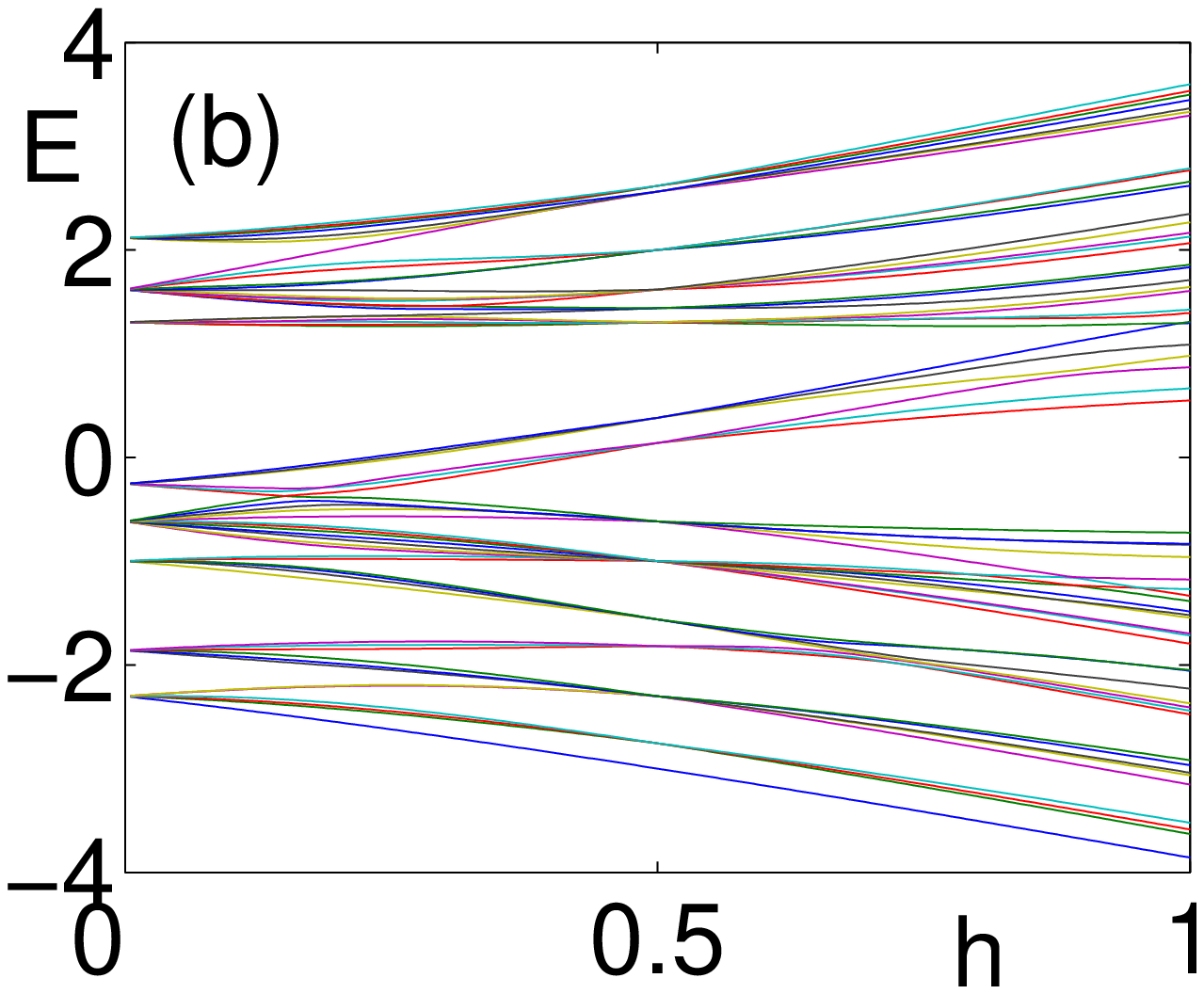}
\includegraphics{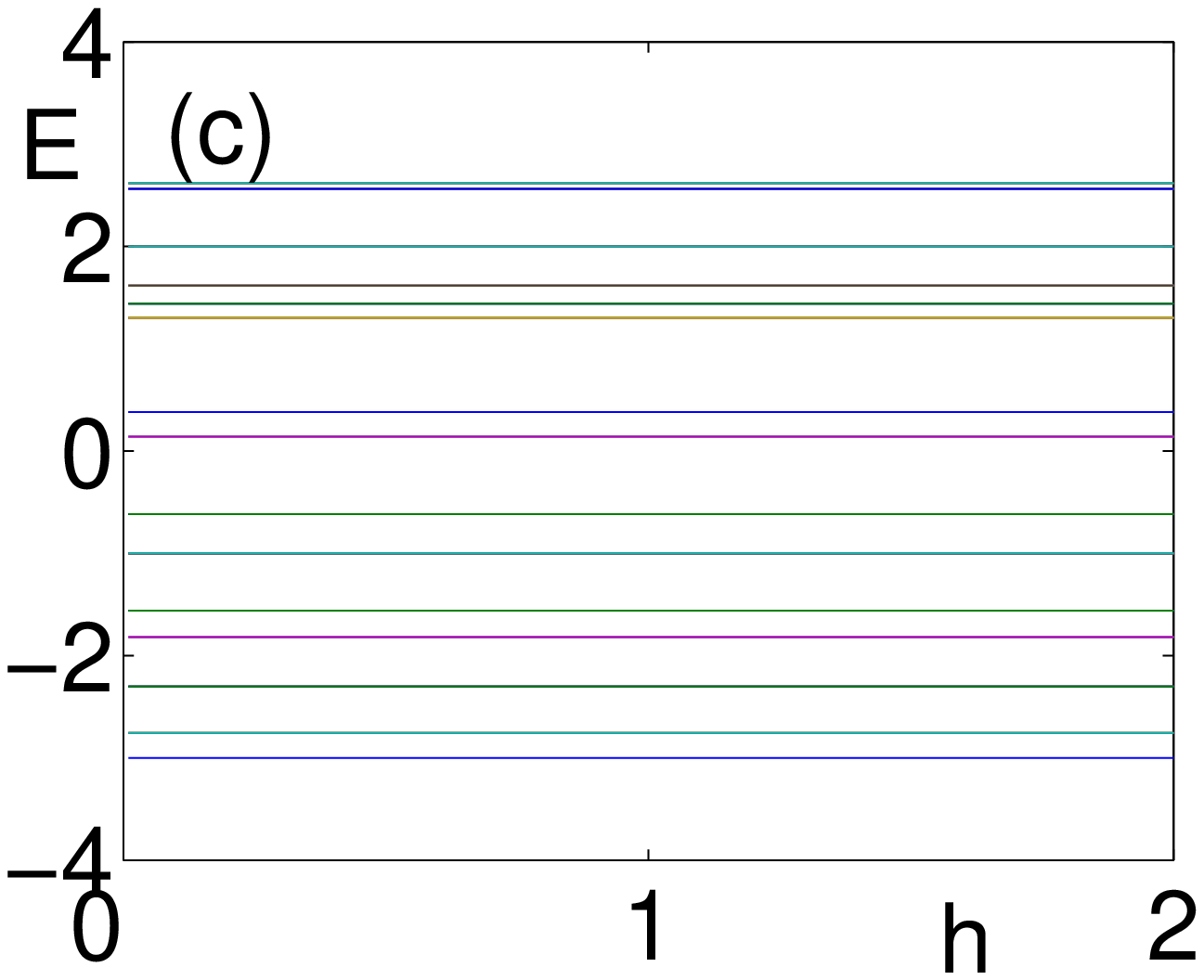}
 }
\caption{\label{C60} The spectrum of the C60 molecule as a function of the
double bond coupling, $h$, with single bond coupling equal to $1$. There are
six modes (two triplets) that are near zero energy. (a) With Kekul\'e
distortion. (b) With cuts between pairs of vortices. (c) With enlarged
vortices, where all double bonds are removed.}
\end{figure}
\end{center}
\begin{center}
\begin{figure}[h]
\resizebox{\linewidth}{!}
{\includegraphics{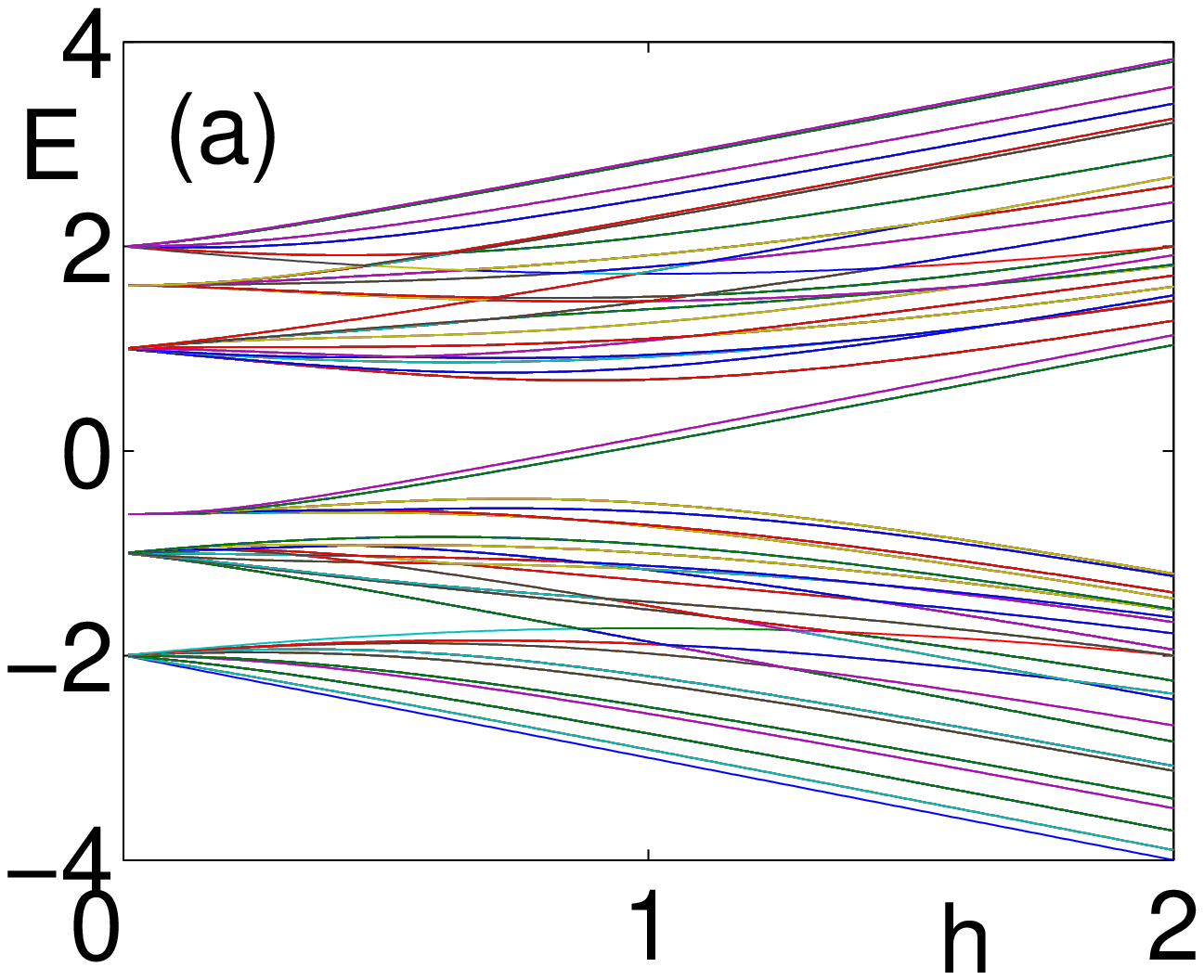}
\includegraphics{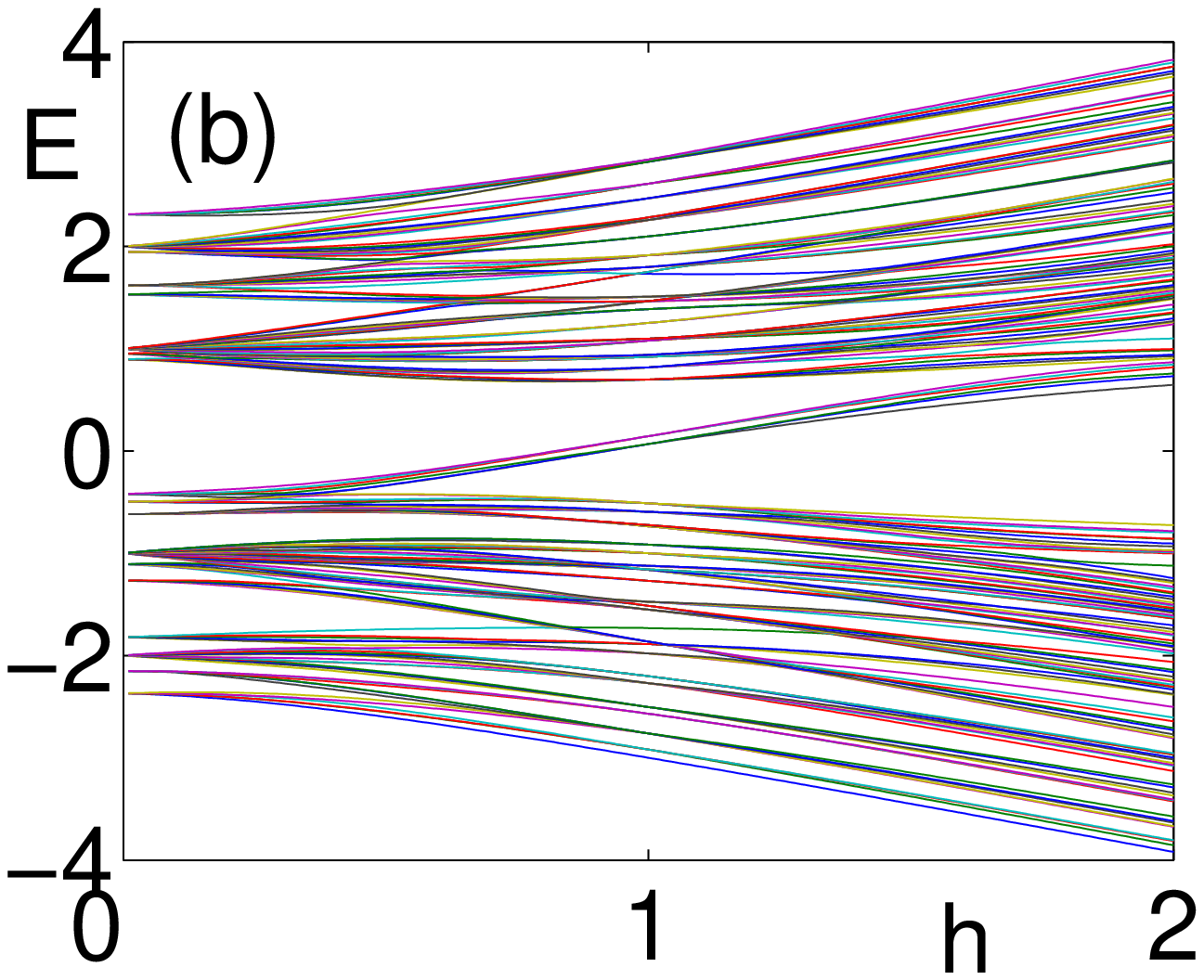}
\includegraphics{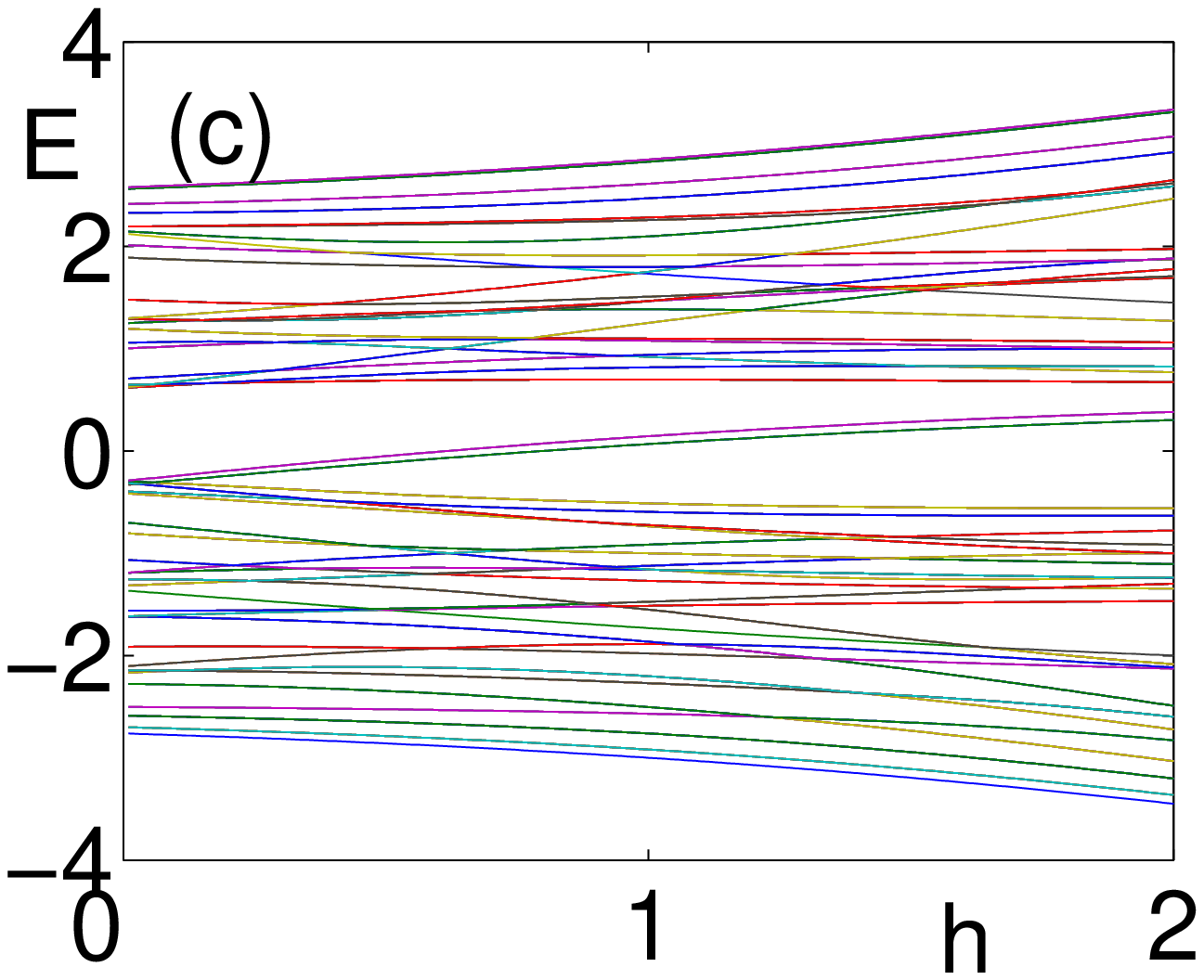}
 }
\caption{\label{C180} The spectrum of the  C180 molecule.
(a) With Kekul\'e distortion. (b) With cuts between pairs of vortices. (c)
With enlarged vortices.}
\end{figure}
\end{center}

We have seen that the introduction of curvature through pentagonal defects
automatically turns Kekul\'e distorted planar graphene into a gauge theory
of the form introduced by Jackiw and Pi. The defects form half-vortices in
the Higgs field corresponding to the Kekul\'e distortion. We can apply the
Jackiw-Rossi-Weinberg index theorem to find a lower bound on the number of
low energy states on the curved surface. When applied to the leapfrog
fullerenes C$_{60+6k}$ we find that there should be six low-lying modes that
are insensitive to the magnitude of the Kekul\'e distortion. The singular
nature of the vortices is a problem however, and numerical investigation of
the leapfrogs C$_{3n\cdot 60}$ with a uniform distortion shows six low lying
modes that {\it do\/} depend on the distortion ~(Figures~\ref{C60}(a) and
\ref{C180}(a)). This is not surprising as in this form the Higgs field is
discontinuous due to its half vorticity at each pentagon. To rectify it one
must allow the vortices to pair up by joining them with cuts, that is
regions with $\Phi=0$. Moreover, the energy contribution from the scalar
field $\Phi$ is given by $\int d^22 r \{|{\bf D}\Phi|^2 +V(\Phi^*\Phi)\}$,
where $V$ is a function with a minimum at $\Phi_0$ with
$V'(\Phi_0^*\Phi_0)=0$ that enforces $\Psi$ to acquire mass. To avoid
divergences of the energy the scalar field~\cite{Jackiw} has to vanish
polynomially as $r\rightarrow 0$. This forces us to consider vortices with
enlarged size. As we shall see in the following these modifications in the
configuration of the Kekul\'e distortion of a fullerene molecule provide a
spectrum that is in agreement with the prediction of the index theorem.

Figures~\ref{C60} and \ref{C180} depict the spectrum of C60 and C180
molecules with Kekul\'e distortion for the various cases of uniform and
smoothed Higgs fields. The coupling of the single bond has been set to $1$
and that of the double bonds is denoted by $h$. The case of uniform Kekul\'e
distortion, presented in Figure~\ref{C60Kekule}(a), has been modified in
order to ensure smoothness of the effective Higgs field. For that we first
introduced cuts among the vortices, i.e. we removed double bonds along paths
that connect pairs of vertices, as seem in Figure~\ref{C60Kekule}(b).
Moreover, we enlarged the size of the vortices by removing double bonds from
the links attached to the pentagons, as seen in Figure~\ref{C60Kekule}(c).
When the last process is performed on the C60 molecule it removes all double
bonds due to its small size.

As expected from the index theorem there are six modes with near zero energy
when the coupling $h$ is varied. While for the case of uniform Kekul\'e
distortion these modes seem to be sensitive to variations of $h$ (see
Figures~\ref{C60}(a) and~\ref{C180}(a)) this is rectified by introducing
cuts and by enlarging the vortices. These processes make the Higgs field
continuous and the energy of the low-lying modes becomes even closer to zero
with an energy that is insensitive to variations of $h$. Indeed, in
Figures~\ref{C60}(c) and~\ref{C180}(c) there are six low-lying modes which
are to a good degree insensitive to variations of $h$. The difference of
their energy from being exactly zero is due to the small size of the system
and it is expected to converge to zero when larger molecules are
employed~\cite{Gonzalez1}.


To summarize, we explicitly demonstrated that the effective gauge field
induced in a graphene sheet, when it is geometrically deformed, couples
non-trivially to the Higgs field induced from a distortion in the tunnelling
couplings. As a result, the fullerene-like molecules have six half vortices
that are {\em fractionally charged}~\cite{Chamon} giving a natural setting
where this topological effect appears. By employing the index theorem we
demonstrated that there should be one zero mode for each pair of such
vortices. While the effective gauge field is responsible for the presence of
the zero modes the scalar field assures that the corresponding degenerate
states are not locally distinguishable implying their topological robustness
(topological degeneracy). In the case of fullerene-like molecules a study of
the corresponding low-lying states showed that a Kekul\'e distortion is
energetically favorable actually reducing the overall energy of the
electrons.

{\em Acknowledgements.} We thank M. Franz, V. Gurarie and R. Jackiw for
inspiring conversations. Work in Urbana was supported by the NSF under grant
DMR-06-03528 and in Leeds by the European Research and Training Network
EMALI and the Royal Society. We would like to thank the Centre for Quantum
Computation at DAMTP Cambridge, and Churchill College Cambridge for
hospitality during the early stage of this project.

\end{document}